\documentclass{llncs}
\usepackage{comment}
\usepackage{url}
\usepackage{algpseudocode}
\usepackage{algorithm}
\usepackage{nth}
\usepackage{amssymb}
\usepackage{epsfig}
\usepackage{threeparttable}
\usepackage{footmisc}
\usepackage{hyperref}
\usepackage{paralist}
\usepackage{wrapfig}
\usepackage{enumitem}
\usepackage{subfig}
\usepackage[table]{xcolor}
\usepackage{color}
\usepackage{multirow}
\usepackage{graphicx}
\newenvironment{packed_enum}{
\begin{enumerate}
  \setlength{\itemsep}{1pt}
  \setlength{\parskip}{0pt}
  \setlength{\parsep}{0pt}
}{\end{enumerate}}

\usepackage{geometry}
\newcommand{\sharma}[2]{\noindent\textcolor{red}{~#1~}\textcolor{red}{\textbf{(Sharma):}~\textcolor{blue}{#2} \\}} 

\begin{document}
\title{\textbf{\textt{CoWiz}:} Covid-19 Analysis and Visualization Dashboard Based On Multilayer Network Analysis}
\title{\textbf{\textt{CoWiz}:} Visualization of Aggregate Analysis of Covid-19 Data Using Multilayer Networks}
\title{\textbf{\textt{CoWiz++}:} Animated Visualization of Covid-19 Aggregate Data Analysis Using Multilayer Networks}
\title{\textbf{\textt{CoWiz++}:} Multiple Visualization Dashboard For Covid-19: Of Base Data And Analysis Using Multilayer Networks}
\title{\textbf{An Extensible Dashboard Architecture For Visualizing Base And Analyzed Data}}

\author{Abhishek Santra, Kunal Samant, Endrit Memeti, Enamul Karim\\ and Sharma Chakravarthy}
\institute{IT Laboratory \& CSE Department, UT Arlington}

\maketitle

\begin{abstract}
\setlength{\parindent}{3ex}

\vspace{-25pt}

Any data analysis, especially the data sets that may be changing often or in real-time, consists of at least three important synchronized components: i) figuring out what to infer (objectives), ii) analysis or computation of objectives, and iii) understanding of the results which may require drill-down and/or visualization. There is a lot of attention paid to the first two of the above components as part of research whereas the understanding as well as deriving actionable decisions is quite tricky. Visualization is an important step towards both understanding (even by non-experts) and inferring the actions that need to be taken. As an example, for Covid-19, knowing regions (say, at the county or state level) that have seen a spike or prone to a spike in cases in the near future may warrant additional actions with respect to gatherings, business opening hours, etc. This paper focuses on an extensible architecture for visualization of base as well as analyzed data.

This paper proposes a modular architecture of a dashboard for user-interaction, visualization management, and complex analysis of base data. The contributions of this paper are: i) extensibility of the architecture providing flexibility to add additional analysis, visualizations, and user interactions without changing the workflow, ii) decoupling of the functional modules to ease and speedup development by different groups, and iii) address efficiency issues for display response time. This paper uses Multilayer Networks (or MLNs) for analysis.

To showcase the above, we present the implementation of a visualization dashboard, termed \textbf{\texttt{CoWiz++}} (for \textbf{\texttt{Co}}vid \textbf{\texttt{Wiz}}ard), and elaborate on how web-based user interaction and display components are interfaced seamlessly with the back end modules.
\end{abstract}

\section{Motivation}
\label{sec:introduction}



\setlength{\parindent}{3ex}

Since early 2020, when the Covid-19 cases were first reported in the US, the virus has spread to all 3141 US counties\footnote{We focus on the USA as we have more accurate data for that although the pandemic is worldwide! Any country can be analyzed by swapping the data sets and with minor changes, such as prefectures in Japan instead of states.} in all states at different rates. As the hunt for a vaccine was launched, the number of cases has grown and leveled off based on the actions taken by different counties and states. Lack of a national policy and lack of synchronization between state and federal mandates have resulted in undesirable situations as compared to other coordinated efforts in other parts of the world. 
From a data collection viewpoint, a number of sources provide features associated with a confirmed report, such as infected case, hospitalization, death, or recovery making this data set complex with diverse entity (person, county), feature (case, hospitalization, vaccination, ...), and relationship (similarity in cases, hospitalizations, vaccinations, ...) types. 

Currently, many visualizations are used to plot the\textit{peak, dip, and moving averages or colored maps} of Covid data, \textbf{without much analysis on the base data or inclusion of associated data}~\cite{JHU-Covid,NYTimes-Covid,CDC-Covid,UW-Covid,Worldometer,WHO-Covid}. In other words, most of these are focused on the visualization of base data using simple statistical computations. However, for a comprehensive understanding of the spread of the pandemic (or any data for that matter), there is a need to \textit{analyse and compare the effects of different events (mask requirement, social distancing, etc.) and demographics, in multiple geographical regions across different time periods}. 

Broadly, visualizations for a data set can be classified into:
\begin{packed_enum}
    \item [I.] \underline{\textit{Visualization  using base data:}} There is no \textit{analysis} involved in this visualization. Visualization includes primarily statistical information. Attributes and visualization alternatives can be selected by the end-user. Temporal ranges, animation, and other visualization parameters can also be chosen.
    \item[II.] \underline{\textit{Visualization using analyzed data:}} There is an \textit{explicit analysis} that needs to be done on base and associated data prior to visualization. Visualization may include analysed results, drilled-down details of results, as well as visualization alternatives for them. 
    Typically a model is used and objectives computed using that model. 
\end{packed_enum}

Currently available online dashboards/visualizations primarily address category I above. For example, JHU (Johns Hopkins University) dashboard \cite{JHU-Covid}
shows a lot of base data and shows some of them also on a US map with circles indicating the numbers to get a relative understanding. Similarly, the the WHO (World Health Organization) dashboard \cite{WHO-Covid}
shows base data for the world and a clickable map to show some base data for that country. For Covid data, most dashboard focus on either reporting and/or visualizing daily cases on maps (\cite{Worldometer,CDC-Covid,WHO-Covid,UV-Covid}) or  time series plots and statistical information (\cite{JHU-Covid,UW-Covid,NYTimes-Covid}). 

However, for category II, there is a need to \textbf{model} the base data which is dependent on the semantics of the data set. As an example, for Covid data, analysis is based on counties/states. We need to model \textit{entities and relationships}  in order to \textit{analyze} and understand the data set from \textit{multiple perspectives}. The result needs to be \textit{visualized} to maximize understanding. In this paper, we use the Covid-19 data set as well as related information, such as population, average per capita income, education level etc. The focus is on an interactive dashboard architecture that is \textbf{modular, flexible, provides good response time, and supports both categories I and II above}.

For the analysis part, this dashboard uses the  widely-popular Entity-Relationship (ER) model and its conversion to \textbf{Multilayer Networks or MLNs}~\cite{ER/KomarSBC20}. MLNs can handle multiple entities, relationships and features. Informally, MLNs
are layers of networks where \textit{each layer is a simple graph and captures the semantics of a (or a subset of) feature of an entity type}. The layers can also be connected. Moreover, an efficient \textbf{divide and conquer based approach} proposed in~\cite{ICCS/SantraBC17,ICDMW/SantraBC17} is used to analyze specified objectives.

The primary contributions of this paper are:

\begin{itemize}
    \item An \textbf{interactive web-based dashboard} for analysis and visualization
    \item A \textbf{modular} architecture to minimize interaction between the modules to facilitate development of the system by multiple groups with different skill sets.
    \item \textbf{Extensibility} of each module to add analysis, visualization, or interaction/display alternatives with minimal effort. 
    \item \textbf{Multiple visualizations} of base and analysis results.
    \item Use of \textbf{multilayer network for modeling} for analysis underneath.
\end{itemize}

This paper is organized as follows. Section~\ref{sec:related-work} discusses related work. Section~\ref{sec:dashboard-architecture} details the architecture of the dashboard in terms of its modules.
Section~\ref{sec:experimental-validation} presents base and objective-based analysis visualizations for the Covid-19 data set. Conclusions are in Section~\ref{sec:conclusions}.

\section{Related Work}
\label{sec:related-work}

Currently available online dashboards address category I and focus on reporting and visualizing daily cases on maps (\cite{Worldometer,CDC-Covid,WHO-Covid,UV-Covid}) or  time series plots and statistical modeling (\cite{JHU-Covid,UW-Covid,NYTimes-Covid}). They are more focused on visualizing the base daily data. In contrast,
drill-down of analysis of results is critical especially for complex data which has both structure and semantics. For example, it is not sufficient to know the identities of objects in a \textit{community} (e.g., similar counties), but also additional details of the objects (e.g., population, per capita income etc.) Similarly, for a \textit{centrality hub} or a \textit{frequent substructure}. As we are using the MLNs as the data model, we also need to know the objects across layers and their inter-connections. From a computation/efficiency perspective, minimal information is used for analysis and the drill-down phase is used to expand upon to the desired extent. Our algorithms, especially the decoupling-based, make it easier to perform drill-down without any additional mappings back and forth for recreating the structure. Our schema generation also separates information needed for drill-down (Relations) and information needed for analysis (MLNs) from the same Enhanced Entity Relationship (EER) diagram. See~\cite{ER/KomarSBC20} for details.

Visualization is not new and there exists a wide variety of tools for visualizing both base data, results, and drilled-down information in multiple ways~\cite{CDC-Covid,JHU-Covid,UW-Covid}. Our focus, in this paper, is to make use of available tools in the best way possible and not propose new ones. For example, we have experimented with a wide variety of tools including, maps, individual graph and community visualization, animation of features in different ways, hovering to highlight data, and real-time data fetching and display,  based on user input from a menu. The main contribution of visualization is our architecture with a common back end to drive different user interaction and visualization front ends. We have also paid attention to efficiency at the back end by caching pre-generated results and use of an efficient data structure for lookup~\cite{ICDE2021Demo/Cowiz}.

\textit{Community detection} algorithms have been extended to MLNs for identifying tightly knit groups of nodes based on different feature combinations (\cite{
xin2018community,magnani2019community}.) 
Algorithms based on matrix factorization \cite{dong2012clustering}, 
cluster expansion~\cite{li2008scalable}, 
Bayesian probabilistic models \cite{xu2012model}, 
regression \cite{cai2005mining}
and spectral optimization of the modularity function based on the supra-adjacency representation \cite{zhang2017modularity}
have been developed. Further, methods have been developed to determine \textit{centrality measures} to identify highly influential entities \cite{
sole2014centrality,zhan2015influence}. 
However, all these approaches \textit{analyze a MLN  by reducing it to a simple graph} either by aggregating all (or a subset of) layers or by considering the entire MLN as a whole, thus leading to loss of semantics as the entity and feature type information is lost.  


\section{Dashboard Architecture for Analysis and Visualization}
\label{sec:dashboard-architecture}

As part of our research on big data analytics (using graphs and Multilayer Networks), we felt the need for drill-down and visualization of results for understanding and ground truth verification. The results of \textit{aggregate analysis} as compared to statistics, requires looking into more details (or drill-down). For example when a community of counties are computed or centrality nodes (cities) are identified, it is important to understand related information such as population density, per capita income, education level, etc. This was further exacerbated by the fact that the data sets we deal with have multiple types of entities, features, and relationships. So, drill-down and visualization of analyzed data along with additional details became pronounced.

When we started analyzing Covid data using our multilayer network analysis, it was important not only to drill-down, but also to visualize the data set and analysis results in multiple ways combining different aspects of the data set. For example, it was useful to visualize newly active cases in multiple states on a daily/weekly basis to see how they were changing. This could be done for multiple features, such as deaths, hospitalizations, etc. We also wanted to visualize similar regions in the country that had same/similar increase/decrease in active cases over the same time period. This would be very useful in understanding the effects of certain measures taken (e.g., masking, lockdown, social distancing) in different parts of the country. This essentially involved processing the same data under the categories I and II indicated above. This is also true for other data sets.

As we tried to develop a dashboard for Covid-19 visualization, we realized that the skill sets needed for analysis was significantly \textit{different} from those needed for visualization/user-interaction. Analysis required a much deeper understanding of the knowledge discovery process including modeling of the data, coming up with objectives and computing them efficiently. On the other hand, visualization required a deeper understanding of the packages that can be used based on what and how we wanted to display. The client module needed yet another different set of skills in terms of layout, menu design, Java Script, and CSS. It seemed natural that these could be developed by different individuals or groups with appropriate skills if the dashboard can be \textit{modularized along these functional components}. This primarily motivated our architecture shown in Figure~\ref{fig:arch}.

The second thing we noticed was that  most of the currently available visualization dashboards seem to be application and analysis specific. That is, if the data set description and application objectives change over a period of time, then the entire system has to be re-built. Although there is likely to be a separation between the client and back end module, having a single back end module seemed to defeat extensibility in addition to modularity. This would create bottlenecks for progress making the development process quite inefficient. So, \textit{extensibility at the module level} requirement was born out of this observation.

Finally, ability to visualize the same data in multiple ways is extremely important from an understanding perspective. For example, one may want to visualize Covid cases/deaths/hospitalizations as a temporally animated graph for different states. One may also want to see the same data to make decisions by comparing geographical regions using MLN analysis. Multiple visualizations and analysis capability in CoWiz++ follows directly from the extensibility aspect of the architecture. Currently, we support two visualization (one from each category above) as part of the visualization management module and  multiple analysis (base and MLN) in the analysis module. We plan on adding more to each category.

\subsection{Components of the Dashboard Architecture}
\label{sec:arch}

\begin{figure}[h]
\centering
\vspace{-20pt}
\includegraphics[width=\columnwidth]{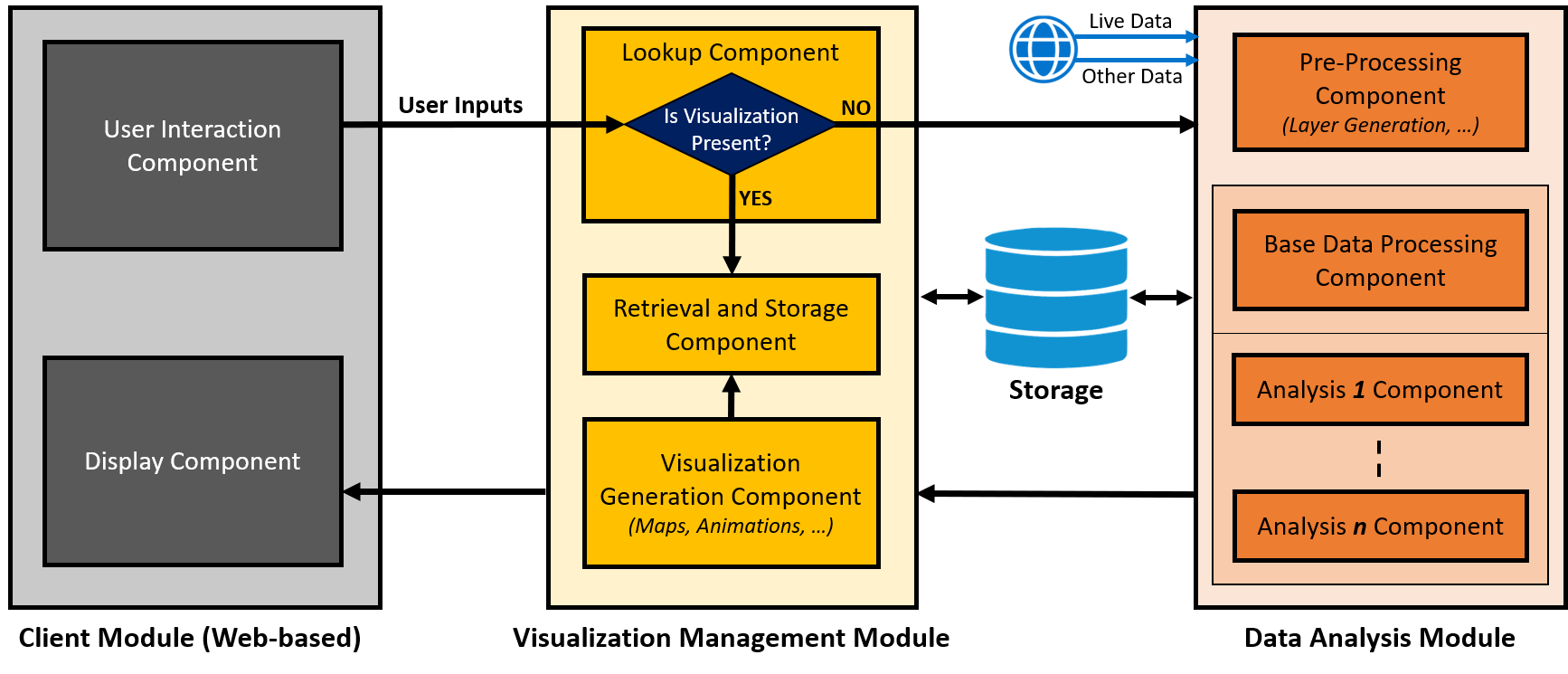}
\vspace{-20pt}
\caption{Modular \texttt{CoWiz++} Dashboard Architecture}
\vspace{-20pt}
\label{fig:arch}
\end{figure}

Our proposed architecture and its components shown in Figure~\ref{fig:arch} has been designed to support the above observations: modularity with \textit{minimal interaction} between the modules and extensibility within \textit{each module}. These two, when incorporated properly,  facilitates re-use of code within each module and the developing modules independently (by different groups) from one another for different applications. This is one of the major contributions of this paper, where we introduce 3 decoupled modules, each optimized for a specific functionality: \textit{i) a web-based client}, \textit{ii) visualization management}, and \textit{iii) data analysis} modules. This architecture permits the  optimization of each component, independently by separate groups with appropriate skill sets resulting in a flexible, extensible and efficient dashboard. In this paper, we show how the different modules interact and how a mix and match of analysis and visualization can be achieved. Also, note the minimal interaction between the modules (mainly parameters, file handles, etc.) As large number of files are used/generated by the two back end modules, a persistent storage is needed to store them.

There is a need for a closer synchronization between the client module and the back end visualization management module. For this to work correctly, the first step was to identify a \underline{web framework} that can support these two modules synergistically. The other considerations were: seamless communication, ease of use, availability of detailed documentation and strong open-source community for future development and extensions. Support for web deployment for increased portability was important.

\begin{table}[h]
    \centering
    \vspace{-15pt}
    \begin{tabular}{|m{1.3cm}|m{2.2cm}|m{1.6cm}|m{2cm}|m{2.5cm}|m{2.2cm}|}
        \hline
         & \textbf{Minimalistic} & \textbf{Language} & \textbf{Plotly Compatibility} & \textbf{Documentation Available} & \textbf{Flexibility and Control}\\
         \hline
         \textcolor{blue}{\textbf{Flask}} & \textcolor{blue}{\textbf{Yes}} & \textcolor{blue}{\textbf{Python}} & \textcolor{blue}{\textbf{Yes}} & \textcolor{blue}{\textbf{Extensive}} & \textcolor{blue}{\textbf{High}} \\
         \hline
         Django & No & Python & Limited & Extensive & Low\\
         \hline
         Vaadin & No & Java & No & Limited & Low \\
         \hline
    \end{tabular}
    \caption{Web Framework Alternative and Feature Comparison}
    \label{table:framework-alts}
    \vspace{-30pt}
\end{table}

Table \ref{table:framework-alts} lists the features of the widely used web frameworks we considered. The \textbf{python-based web framework Flask} was chosen over Django and Vaadin, mainly due to its minimalistic, interactive, flexible and extensible characteristics. Flask satisfied all our requirements as shown in the table. Moreover, visualization tools like Plotly are supported exhaustively by Flask, which is not supported by others. Most importantly, as compared to others, it gives maximum flexibility and control due to granular tuning for customisation and makes no assumptions about how data is stored, thus becoming a viable choice for a wider spectrum of applications. 
Below, we describe each module emphasizing the modularity and extensibility aspects.

\noindent\underline{\textbf{1. Client Module (User-Interaction and Display Components):}}
\label{sec:client-module}
Each analysis and visualization uses a set of inputs given by the user for that specific analysis and visualization. The client module is responsible for presenting an unambiguous, clear, and simple user interface for collecting those parameters. Once the parameters and display types are identified, this module can be implemented independently and the collected parameters are passed. The inputs can be in the form of ranges (dates, latitude-longitude, times, ...), lists and sets (features, items, ...) or filtering options. The various elements of this component are supported using HTML and CSS. 

The other task of this module is to display the visualization generated for the input parameters by the other two modules. This is typically in the form of an html file that is displayed using the \textit{iframe component}. In addition to displaying the html file, this component is also responsible for tickers and other relevant information (part of display type.) For example, the visualization of top 10 Covid-19 news articles and the latest cumulative number of cases and deaths is achieved through \textit{scrollable or moving ticker}s, implemented using \textit{JavaScript and AJAX scripts} and the \textit{marquee component}. Note, this is based on the input and is done in real-time.

\begin{wrapfigure}{l}{0.49\linewidth}
\centering
\includegraphics[width=\linewidth]{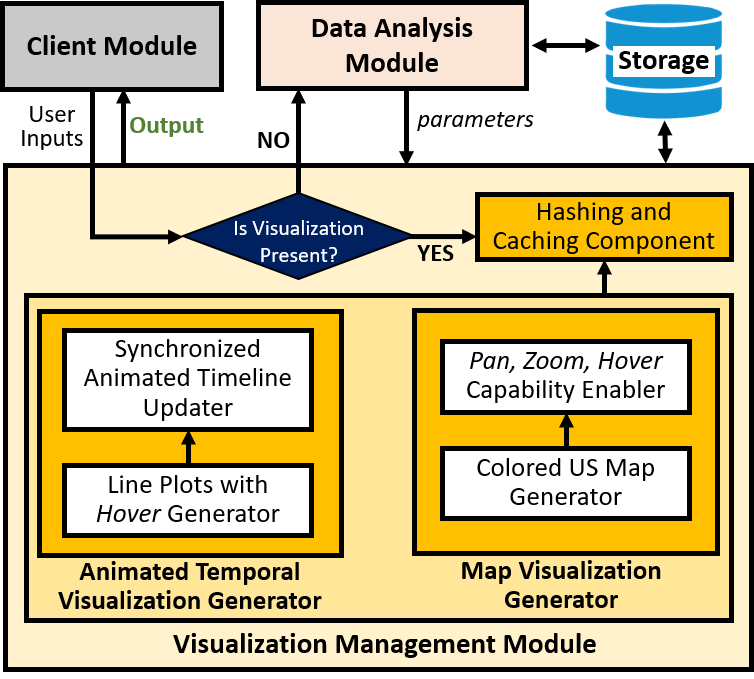}
\caption{Details of the Visualization Management Module}
\vspace{-20pt}
\label{fig:viz-flow}
\end{wrapfigure}

\noindent \underline{\textbf{2. Visualization Management Module (Dashboard Back-End):}}
\label{sec:arch-viz}
Functionally, this is an important module, detailed in Figure~\ref{fig:viz-flow}, that handles several tasks: i) visualization generation -- using either base data or computed results -- from the analysis module, ii) caching the visualization information using an efficient data structure\footnote{Currently, an in-memory hash table is used for quick lookup. If it cannot be held in memory, this can be changed to a disk-based alternative (extendible hash or B+ tree) \textbf{without affecting any other module}. For disk-based, pre-fetching and/or other buffer management strategies can be used to improve response time. Separate hash tables are used for different visualizations for scalability. Also, hash tables are written as binary objects and reloaded avoiding re-construction time.}, and iii) looking up whether the visualization exists for a given set of parameters and display type to avoid re-generation and speedup response-time. As can be seen in Figure~\ref{fig:viz-flow}, there are two separate visualization generation components, a hash and cache component for quick lookup and storage. Additional visualization generation modules can be easily added. This module interacts with the other two modules and the storage.

Since analysis and generation of the visualization accounts for most of the response time, we have used two known techniques for improving response time: \textit{i) materialization of previous analysis results} -- widely used in DBMSs to trade off computation with space and \textit{ii) efficient hash-based lookup} to identify whether a materialized visualization exists. 
The first check in the this module is to find the presence of the display file generated earlier.  As there are hundreds of thousands of possible user input combinations, avoiding collisions in hashing is important. If the display is present, it is used. If not, the parameters are sent to the analysis module to generate computed results so this module can generate the visualization after that. \textbf{This approach has shown significant improvement and has reduced the response time from 15 seconds to 3 seconds (400\% improvement)} for map visualizations. This module uses packages from Python, R and Tableau to provide diverse types of interactive graphical visualizations. Two visualizations are currently supported by this module and are discussed briefly below. Also, note that the display as well as the ticker information is based on user input. Currently, after multiple user interactions from 10+ countries, 3000+ community result files and 400+ map visualization files are present.


\noindent \textbf{Animated Temporal Visualizations}: This is an example of visualization used for category I discussed earlier. For comparing the variation in 2 selected features for up to 5 states, \textit{2 side-by-side synchronized animated timeline plots} are generated, where the per day (or per period) values are plotted for each feature corresponding to the states selected. Visualization is generated using Python's popular \textit{Plotly library and R language}, where, in each plot, the y-axis corresponds to one of the feature's values and x-axis to the timeline. The two plots are displayed side-by-side in a single plotly subplot, to implement the \textit{synchronised timeline with play and pause features}. 

The \textit{play and pause features} are implemented in the plotly subplot as a slider. When user clicks on \textit{play} button, the plotting of data points on both charts \textit{begins} from the first available day. The position of the \textit{slider tick} corresponds to the last day for which data is plotted. Clicking on a tick constitutes a \textit{pause}. A \textit{legend and hover feature} are implemented on these plots to  visualize the details. By \textit{double-clicking on a state} on the \textit{legend}, \textit{only} its corresponding line curves are made visible on the 2 plots. The plots can be repopulated by clicking the other legend items. The \textit{hover feature} is also incorporated to display the \textit{feature value of a state} for the \textit{date in focus}. Finally, this animation is saved as an \textit{html file}. These visualizations and their objectives are shown in Figures~\ref{fig:vacc-trips} and ~\ref{fig:west-virginia} in Section~\ref{sec:experimental-validation}.


\textbf{Map-Based Visualizations}: Communities are generated as part of the category II analysis objectives ~\ref{C3-3} and \ref{C2-1}, 
where counties are clustered based on similar change in a feature (in this case, similar change in \textit{new cases}). Each county in the community is displayed on a colored US map based on the \textit{severity} of changes in Covid cases reported in its assigned community which corresponds to a range - from SPIKE (as {\color{red}\textbf{red}}) to BIG DIP (as {\color{green!30!black}\textbf{green}}). The FIPS (Federal Processing Information Standards) codes of the US counties present in the generated community file are used by the \textit{choropleth\_mapbox() function of Python's plotly} library to generate colored counties on the US map with \textit{pan and zoom capability} enabled. Moreover, the census information available as part of this file is used to generate the \textit{hover text} for counties. The generated US map for a community file is stored as an \textit{html file}. This visualization for objectives \ref{C3-3} and \ref{C2-1} is shown in Figures~\ref{fig:springbreak} and ~\ref{fig:vaccine} in Section~\ref{sec:experimental-validation}\footnote{Dashboard: \url{https://itlab.uta.edu/cowiz/}, Youtube Video: \url{https://www.youtube.com/watch?v=4vJ56FYBSCg}. \textbf{Reviewers are encouraged to play with the dashboard and watch the video.}}.


\noindent\underline{\textbf{3. Data Analysis Module (Dashboard Back-End)}}
\label{sec:arch-analysis}
The analysis module is another key module of the architecture. It supports several components that are important for different aspects of data analysis: \textit{i) extraction of relevant data} from external sources, \textit{ii) pre-processing} of extracted (or downloaded) data, and  \textit{iii) generate results} for both base and analysis alternatives. It is more or less agnostic to visualization except to generate information needed for visualization, but  does not even know how they are visualized. All three components are extensible in their own right and only rely on the user input passed from client module through the visualization management module. This modules interacts with the persistent storage for both input data and output generated. This module generates output for visualization. Of course, base data preparation is much  simpler than the other one.

\textbf{Extraction/Downloading and Pre-processing Component}: Components of this module are responsible for the real-time extraction of data from identified web sources (e.g., New York Times, WHO, CDC), update of data that changes periodically -- once a day/week/month (e.g., Covid data in our dashboard) using \textit{cron jobs}. For example, when date ranges are input by the user as part of the menu, that information is used to extract information \textit{only for those periods}, pre-processed (cleaned, filtered, sorted/ranked), and prepared for visualization. All pre-processing of data extracted from real-time sources as well as base data used for analysis are done in this component. Examples from the current dashboard are -  \textbf{(i) Period Specific Top 10 Covid-19 Articles}: For category II, 2 periods are provided. From the set of New York Times articles, a subset of top 10 most relevant Covid-19 news articles for the \textit{latest} period specified by the user are filtered using keywords, sorted in reverse chronological order and the top-k are chosen for display, and \textbf{(ii) Latest Cumulative Case and Death Count for Selected States, US and World}: For category I, 
the latest total number of cases and deaths for the user specified states (along with, US and World) are filtered out from the consistent clean WHO and CDC extractions.

\begin{wrapfigure}{r}{0.34\linewidth}
\centering
\vspace{-20pt}
\includegraphics[width=\linewidth]{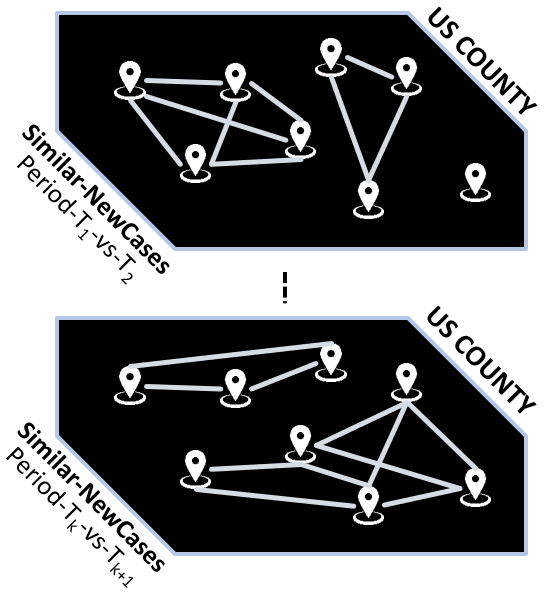}
\vspace{-20pt}
\caption{Multilayer Network for Category II Objectives}
\vspace{-30pt}
\label{fig:covid-homln}
\end{wrapfigure}

\textbf{Complex Analysis Component}: Any analysis for the two categories discussed earlier is  supported by this module. It can be as simple as plotting the base data or generating moving averages to be plotted or a combination thereof. Here, we will focus on category II as it involves modeling, computation of objectives, and drill-down before visualization. Consider understanding the effect of long weekends and vaccination on Covid cases in the US. This can be formulated as objectives to be analyzed on the Covid data set using the MLN model\protect{\footnote{Any analysis approach and associated model can be used. We are using the MLN model proposed in~\cite{ICCS/SantraBC17,Arxiv/SantraKBC20} for this dashboard. This also validates our assertion of the applicability of MLN model  for complex analysis of real-world applications.}}.


   \begin{enumerate}[label={\textbf{(A\arabic*)}}]
        \item \label{C3-3} Which regions got significantly affected due to major long holidays (like Spring Break, Labor Day, Thanksgiving, New Year Celebration, ...)? What precautions need to be taken for future events? The inverse can be computed which may be very helpful as well.
 
\end{enumerate}


\begin{enumerate}[label={\textbf{(A\arabic*)}}, resume]
        \item \label{C2-1} In which regions, was vaccination most effective? That is, how have geographical regions with maximum (and minimum) rise in cases shifted between the periods pre and post the beginning of the vaccination drive?
\end{enumerate}

For \ref{C3-3} and \ref{C2-1}, \textit{geographical regions} need to be analyzed across two periods for similar Covid spread. MLN layers are created using US counties as nodes and connecting them if the change of feature (e.g., new cases (shown in Figure \ref{fig:covid-homln}), deaths, hospitalizations etc.) across the two periods is similar (using slabs of percentages.) Community detection algorithms (e.g., Louvain, Infomap, etc.) on the generated individual MLN layers for detecting communities that will correspond to geographical regions showing similar change in the feature. Any user-selected feature can be used for this purpose. The communities generated are categorized based on the severity of change in covid cases - from spike in cases ($>$100\% increase) to big dip (100\% decrease). This generated community allocation file is enriched by adding the US census data like \textit{population density per sq. mile, median household and percentage of high school} graduates for each county. The results of \ref{C3-3} and ~\ref{C2-1} are shown in Section~\ref{sec:experimental-validation}.

The analysis module only needs to know the information used by the visualization management module and not the actual visualization \textit{type}. This information is known to the client module for each type of user interaction and is passed onto other modules.

\section{User-Interaction and  Visualization  of COVID-19 Data}
\label{sec:experimental-validation}

\begin{figure}[h]
   \vspace{-25pt}
   \centering
   \includegraphics[width=\linewidth]{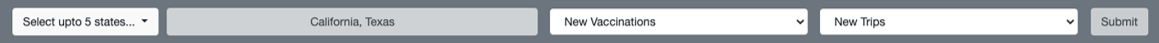}
   \vspace{-15pt}
   \caption{Input Panel for Category I Objectives \textit{(with inputs for Fig. \ref{fig:vacc-trips} visualization)}}
   \label{fig:front-cat1}
   \vspace{-25pt}
\end{figure}  

 \begin{wrapfigure}{l}{0.2\linewidth}
   \vspace{-25pt}
   \centering
   \includegraphics[width=\linewidth]{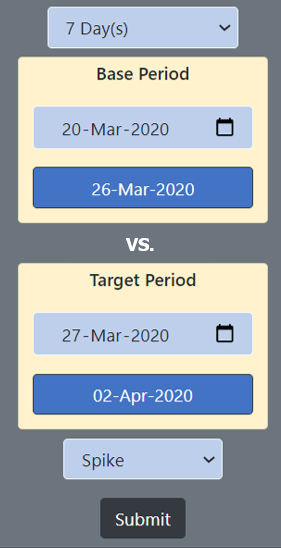}
   \caption{Input Panel for Category II Objectives \textit{(with inputs for Fig. \ref{fig:springbreak} (b) visualization)}}
   \label{fig:front-cat2}
   \vspace{-25pt}
\end{wrapfigure}  

\noindent The \texttt{CoWiZ++} dashboard is hosted on a Apache HTTP Server 2.4.7 on Linux machine. It is supported on all major web browsers. For the best user experience, screen sizes above 1200 pixels are recommended. Figures \ref{fig:front-cat1} and  \ref{fig:front-cat2} show sample user-interaction screens (with input) for two different analyses and visualizations currently supported.

\begin{figure}[h]
   \centering
   \includegraphics[width=0.95\linewidth]{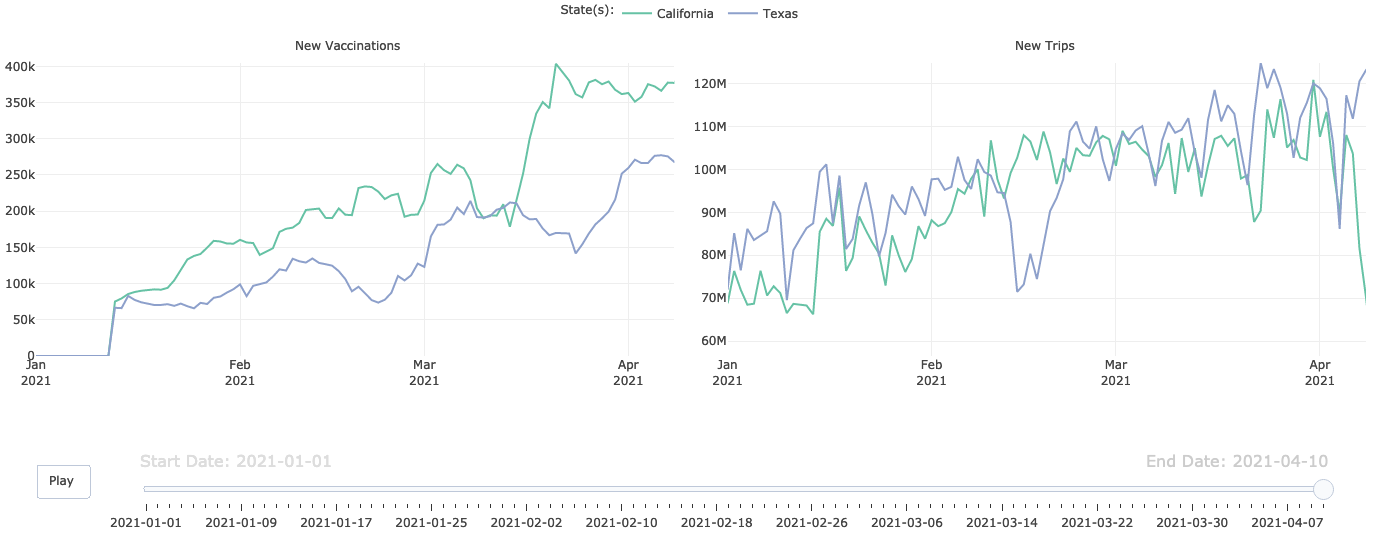}
      \vspace{-5pt}

   \caption{Vaccinations vs. Road Travel Trend in 2 Populous States}
   \label{fig:vacc-trips}
\end{figure}

\noindent \underline {Vaccination vs. Driving:} For \textit{understanding correlation between vaccination and people taking road trips outside their homes}, the user-interaction component shown in Figure \ref{fig:front-cat1} is used. Figure \ref{fig:vacc-trips} shows the completely rendered \textit{snapshot} of the animated timeline depicting the correspondence between the  number of new vaccinations and number of new trips undertaken by people in two of the \textit{largest states by population density} - \textbf{California and Texas}. The plots reveal something interesting. In Texas, new trips have risen disproportionately to the vaccine whereas in California, that is not the case. This conforms to our understanding of the way these two states have handled Covid. Note the difference in scale between the two animations.

 \begin{figure}[h]
   \vspace{-10pt}
   \centering
   \includegraphics[width=0.95\linewidth]{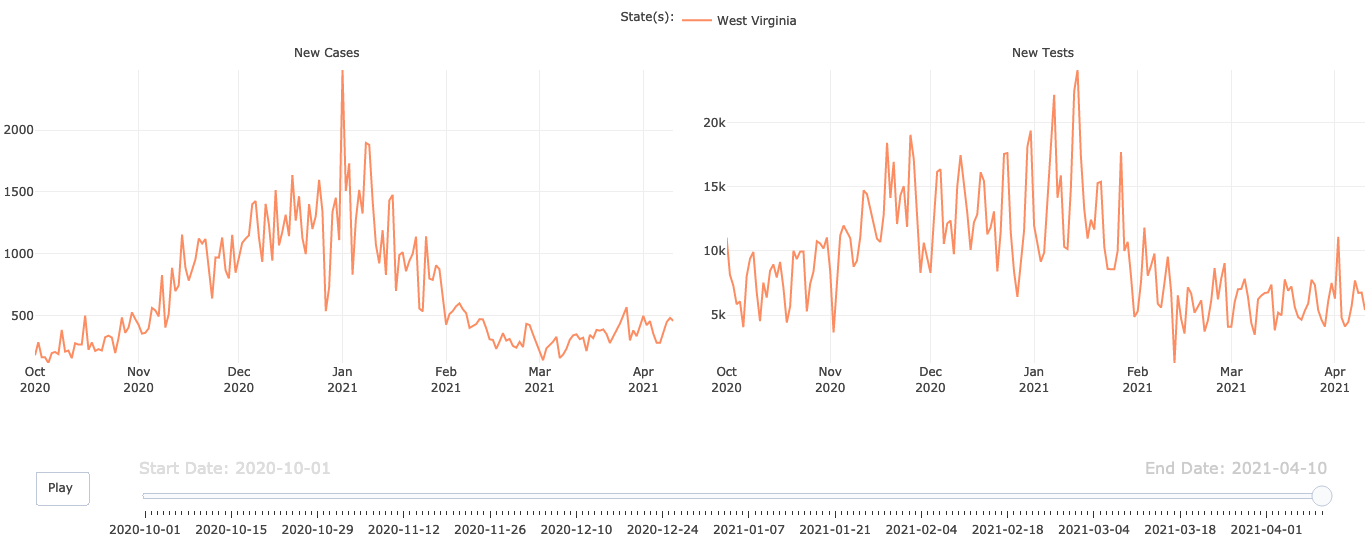}
   \vspace{-5pt}
   \caption{New Cases vs. New Tests in Low Per Capita Income State}
   \label{fig:west-virginia}
   \vspace{-20pt}
\end{figure}

\noindent \underline {New Covid Cases vs. Testing:} Testing for Covid is important according to CDC and should be continued independently of the new cases. For understanding whether this is the case, we considered West Virginia for the paper (please try with other states), one of the low per capita income states.  
Figure \ref{fig:west-virginia} shows these animated plots side-by-side. For an unknown reason, testing seems to follow new cases instead of staying constant. This seems to give the impression that the ones that are being tested are mainly the ones coming with symptoms where as general population is not likely being tested. For a state-level decision maker, this can be useful as an important piece of information discovered through the visualization tool. 

\begin{figure}[h]
   \centering
   \includegraphics[width=0.95\linewidth]{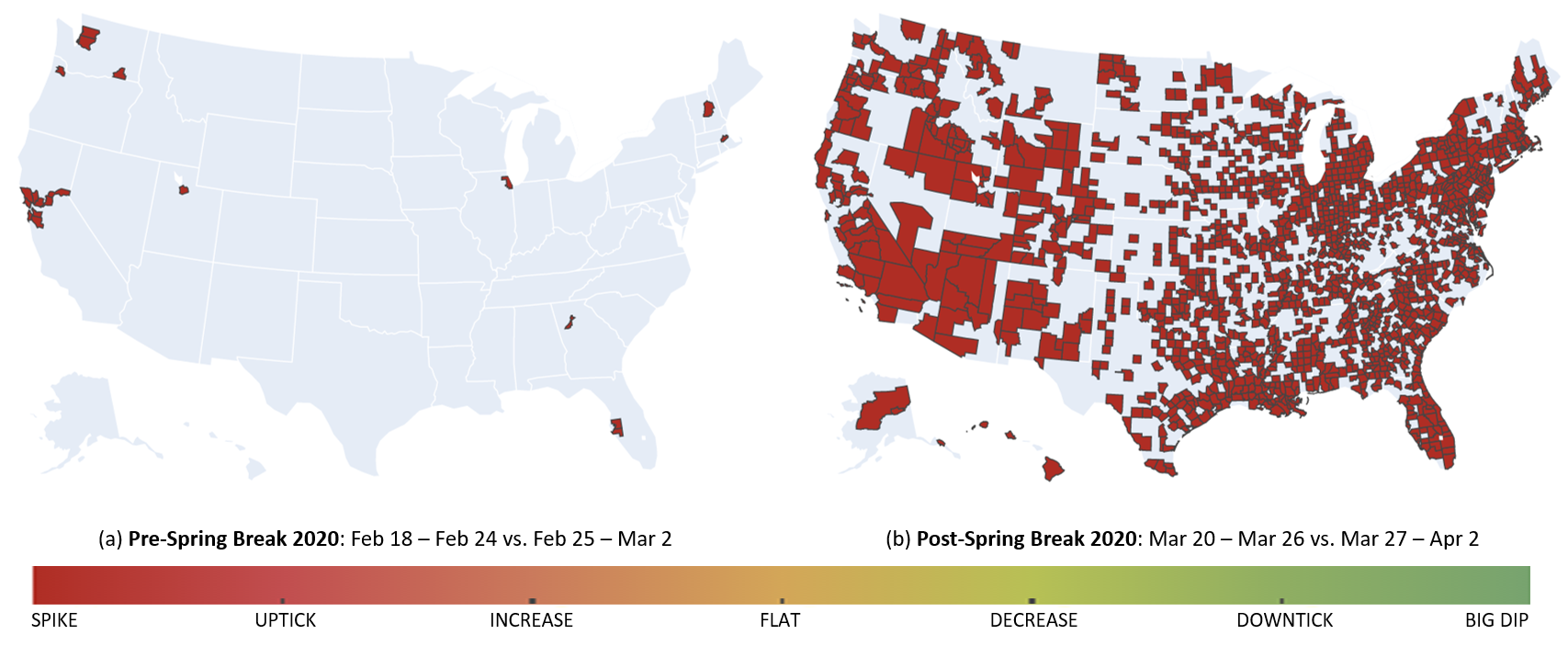}
   \vspace{-10pt}
   \caption{\ref{C3-3}
\textbf{\color{red}SPIKE} in cases due to the 2020 Spring Break}
   \label{fig:springbreak}
   \vspace{-5pt}
\end{figure}

\noindent \underline{Spring Break Effect \ref{C3-3}:} 
For this category II visualization, we use the dashboard front-end shown in Figure \ref{fig:front-cat2}) using the geographical regions with rise in daily confirmed cases in the \textit{pre 2020 spring break} consecutive 7-day intervals - Feb 18 to Feb 24 and Feb 25 to Mar 2. Similar consecutive 7-day periods were chosen for \textit{post 2020 spring break} - Mar 20 to Mar 26 and Mar 27 to Apr 2. This choice is based on the observation that for most US counties, the spring break lasted until the third week of March in 2020. 
The drill-down results have been visualized in Figure \ref{fig:springbreak} that show how \textbf{post the spring break there was a spike in the number of daily cases in counties across the US as compared to pre spring break.} Various reports attributed this massive surge due to the widespread travel to popular tourist destinations during the break leading to \textbf{crowds and non-adherence to social distancing norms}~\cite{springbreak2020-3}. 

\begin{figure}[h]
   \centering
   \includegraphics[width=0.95\linewidth]{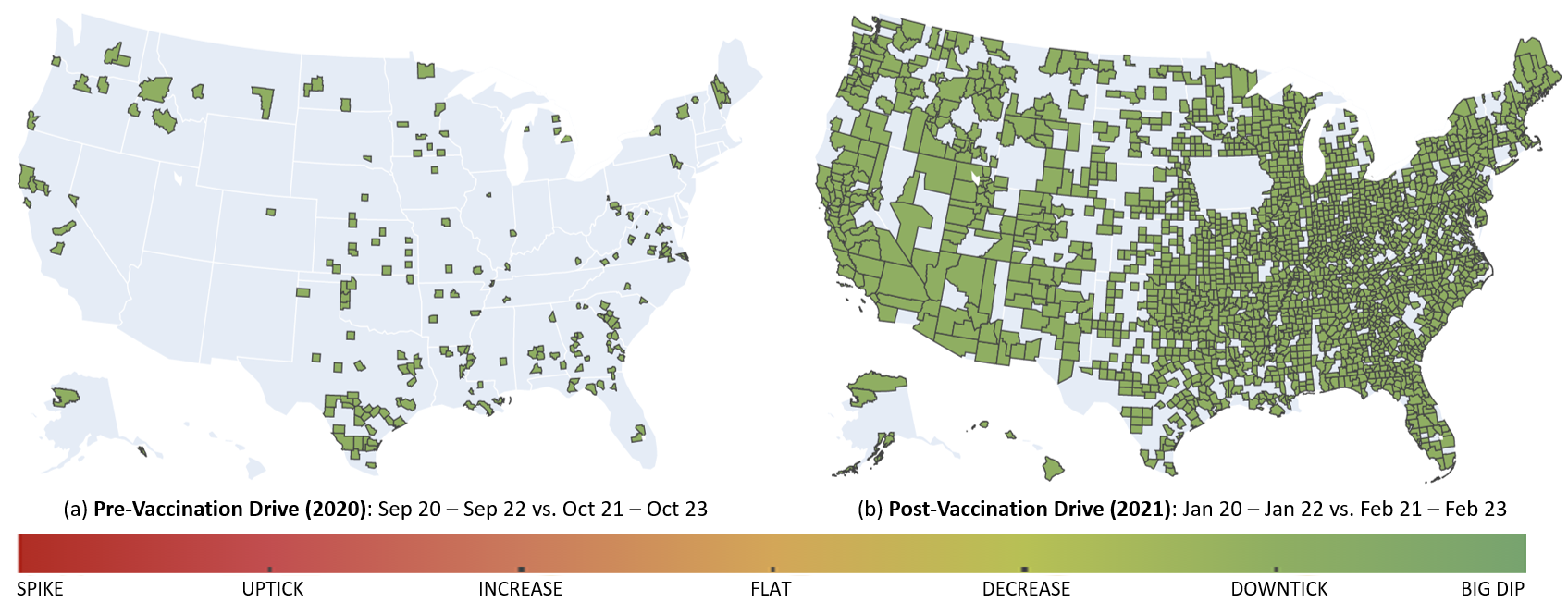}
      \vspace{-10pt}
   \caption{\ref{C2-1} \textcolor{green}{BIG  DIP} due to Vaccination Drive in the US}
   \label{fig:vaccine}
   \vspace{-15pt}
\end{figure}

\noindent \underline{Vaccination Drive Effect\ref{C2-1}:} Here we visualize how the geographical regions with decline in daily confirmed cases shift in month-apart 3-day periods pre and post the \textit{Vaccination Drive}. The vaccination drive in the US began from December 14, 2020 \cite{vaccine-0}. For the \textit{pre} vaccination drive layer, the 3-day intervals considered were Sep 20 to Sep 22 and Oct 21 to Oct 23 in 2020. For the \textit{post} vaccination drive layer, the 3-day intervals were Jan 20 to Jan22 and Feb 21 to Feb23 in 2021. The \textit{community} (groups of counties) results have been drilled-down from the individual layers and the ones displaying a downward trend have been visualized in Figure \ref{fig:vaccine}. This visualization clearly shows how the \textbf{vaccination drive has become one of the reasons that has led to controlling the spread of COVID across US in the past few months}. This fact is also verified from independent sources that say how the administration of the vaccine has led to a \textit{decline in severe cases, hospitalizations and deaths} not only in the US~\cite{vaccine-1,vaccine-2}.

\section{Conclusions}
\label{sec:conclusions}

In this paper, we have presented a modular dashboard architecture to visualize base data, input-based extraction of real-time information, and complex \textit{analysis} results meaningfully. The architecture modules based on functionality provide \textit{flexibility} (of development), \textit{extensibility} (of visualizations, analysis, and data sets), and \textit{efficiency} (for response time). 
Each component within a module is  parameterized making it easier to replace data sets for similar visualization or change visualization for same data set. 

Future work includes adding additional base data, other analysis options, and hierarchical visualizations for country and further into states. Other extensions to support multiple users at the same time and good throughput for large number of users are underway. 


\begin{scriptsize}
\bibliographystyle{plain}
\bibliography{./bibliography/somu_research,./bibliography/itlabPublications,./bibliography/itlabTheses,./bibliography/graph-search,./bibliography/graph-partitioning,./bibliography/JayBib,./bibliography/santraResearch}
\end{scriptsize}
\end{document}